\documentclass[pre,aps,twocolumn,showpacs,floatfix,superscriptaddress]{revtex4}
\usepackage{graphicx} 
\newcommand {\ds}{\displaystyle}
\begin{document}
%\draft
%\tighten
\title{Dynamics and stability of vortex-antivortex fronts in
  type II superconductors}  
\author{C. Baggio} \affiliation{Instituut-Lorentz, Leiden University,
PO Box 9506, 2300 RA Leiden, The Netherlands}

\author{M. Howard} \affiliation{Instituut-Lorentz, Leiden University,
PO Box 9506, 2300 RA Leiden, The Netherlands}\affiliation{Department
of Mathematics, Imperial 
College London, South Kensington Campus, London SW7 2AZ, U.K.}

\author{W. van Saarloos}
\affiliation{Instituut-Lorentz, Leiden University,
PO Box 9506, 2300 RA Leiden, The Netherlands}
\date{\today}
%\maketitle

\begin{abstract}
\noindent
The dynamics of vortices in type II superconductors exhibit a variety
of patterns whose origin is poorly understood. This is partly due to
the nonlinearity of the vortex mobility which gives rise to singular 
behavior in the vortex densities.
Such singular behavior complicates the application
of standard linear stability analysis. In this paper, as a first step
towards dealing with these dynamical phenomena, we analyze the
dynamical stability of a front between vortices and antivortices. In
particular we focus on the question of whether an instability of the
vortex front can occur in the absence of a coupling to the
temperature.  Borrowing ideas developed for singular bacterial growth
fronts, we perform an explicit linear stability analysis which shows
that, for sufficiently large front velocities and in the absence of
coupling to the temperature, such vortex fronts are stable even in the
presence of in-plane anisotropy.  This result differs from previous
conclusions drawn on the basis of approximate calculations for
stationary fronts. As our method extends to more complicated models,
which could include coupling to the temperature or to other fields, it
provides the basis for a more systematic stability analysis of
nonlinear vortex front dynamics.
\end{abstract}
\pacs{74.25.Qt,05.45.-a}

%\begin{multicols}{2}
%\narrowtext
%\setlength{\textfloatsep}{10cm}
\maketitle
\section{Introduction}
{\label{uno}
\subsection{Motivation}

The properties of type II superconductors have been studied
extensively in past decades.  The analysis of patterns in the
magnetic flux distribution has generally focused on equilibrium vortex
phases. The interplay of pinning and fluctuation effects, especially
in the high-$T_c$ superconductors, gives rise to a rich variety of
phases whose main features are by now rather well understood
\cite{giannirmp,gianni}. In comparison with equilibrium behavior,
however, our understanding of the dynamics of vortices, and the
dynamical formation of vortex patterns, is still much less well
developed.
  
Recently, experiments with magneto-optical techniques on flux
penetration in thin films have revealed the formation of a wide
variety of instabilities. An example is the nucleation of
dendrite-like patterns in $Nb$ and $MgB_2$ films
\cite{dendrites,Shantsev,Shantsev2}. These complex structures consist of
alternating low and high vortex density regions and are found in a
certain temperature window. Likewise, flux penetration in the form of
droplets separating areas of different densities of vortices has been
observed in $NbSe_{2}$ \cite{droplets}.  Patterns with branchlike
structures have been found also in high-$T_c$ materials, like
$YBa_{2}Cu_{3}O_{7-x}$ \cite{leiderer}.  In addition, the scaling of
the fluctuations of a (stable) vortex front penetrating a thin sample
has been studied \cite{surdeanu}.

Usually the occurrence of dendrite-like patterns in interfacial growth
phenomena can be attributed to a diffusion-driven, long-wavelength
instability of a straight front, similar to the Mullins-Sekerka
instability \cite{Mullins-Sekerka} found in crystal growth.  In this
paper we therefore investigate the stability of a straight front of
vortices and antivortices which propagate into a type II
superconductor. Furthermore, according to the experimental data
\cite{Frello,Indenbom,Koblischka,Vlasko}, 
the boundary between vortices and antivortices exhibits many features 
suggestive of a long-wavelength instability.

The nucleation of dendrites associated with the propagation of a flux
front into a virgin sample has been attributed to such an interfacial
instability. This results from a thermo-magnetic coupling
\cite{Mints,Shantsev,Shantsev2,Aranson} where a higher temperature leads to a
higher mobility, enhanced flux flow, and hence a larger heat
generation. However, the cause of the instability at the boundary
between fluxes of opposite sign is still being debated. Shapiro and
co-workers \cite{Shapiro} attribute these patterns to a coupling to
the temperature field via the heat generated by the
annihilation of vortices with antivortices.
 On the other hand, Fisher {\em et al.}
\cite{Fisher,Fisher1} claim that an in-plane anisotropy of the vortex
mobility is sufficient to generate an instability.

There are several reasons to carefully reinvestigate the idea of an
anisotropy-induced instability of propagating vortex-antivortex
fronts. First of all, even though this mechanism was claimed to be
relevant for the "turbulent" behavior at the boundaries of opposite
flux regions, the critical anisotropy coefficients found on the basis
of an approximation \cite{Fisher,Fisher1} correspond to an anisotropy
too high to describe a realistic situation, even when a nonlinear
relation between the current and the electric field was considered.
\cite{anisocoeff1,anisocoeff2,anisocoeff3}. Secondly, the
calculation was effectively done for a symmetric stationary interface,
rather than a moving one.  Thirdly, the physical picture that has been
advanced \cite{Fisher} for the anisotropy-induced instability is that
of a shear-induced Kelvin-Helmholtz instability, familiar from the
theory of fluid interfaces \cite{kelvin}. However, it is not clear how
far the analogy with the Kelvin-Helmholtz instability actually extends.

In order to try to settle the mechanism that underlines such phenomena, 
we investigate here the linear stability of the interface between
vortices and antivortices without any approximations in the case where
the front of vortices propagates with a finite velocity.
We perform an explicit linear stability analysis which shows that, in
the presence of an in-plane anisotropy, vortex fronts with sufficiently
large speed {\em are} stable in the absence of coupling to the
temperature.  We shall see that the issue of the stability of fronts
between vortices and antivortices is surprisingly subtle and rich:
while we confirm the finding of Fisher {\em et al.}
\cite{Fisher,Fisher1} that stationary fronts have an instability to a
modulated state, our moving fronts are found to be stable for all
anisotropies. Moreover, our calculations indicate that the stability
of such fronts depends very sensitively on the distribution of
antivortices in the domain into which the front propagates, so it is
difficult to draw general conclusions. 
 
Besides the intrinsic motivation to understand this anisotropy issue,
there is a second important motivation for this work. Our
coarse-grained dynamics of the vortex densities is reminiscent of
reaction-diffusion equations with nonlinear diffusion. This makes the
coarse-grained vortex dynamics very different from the Gaussian
diffusive dynamics of a linear diffusion equation. For example, the
fact that vortices penetrate a sample with linear density profiles 
\cite{bean} is an immediate consequence of this.  More fundamentally, the
dynamically relevant fronts in such equations with nonlinear diffusion
are usually associated with nonanalytic (singular) behavior of the
vortex densities --- such singular behavior has been studied in depth
for the so-called porous medium equation
\cite{porous1,porous2,porous3}, which has a similar nonlinear
diffusion. In the case we will study, the front corresponds to a line
on one side of which one of the vortex densities is nonzero, while on
the other side it vanishes identically. In the regime on which we will
concentrate, this vortex density vanishes linearly near the singular
line. But for other cases encountered in the literature
\cite{Fisher1,otherexp} even more complicated non-linear dynamical
equations arise that are reminiscent of reaction-diffusion type models
in other physical systems. The case of bacterial growth models
\cite{benjacob,Judith} illustrates that the non-linearity of the
diffusion process can have a dramatic effect on the front stability,
so a careful analysis is called for. Nevertheless, in our case
nonlinear diffusion by itself does not lead to an instability of the
front, unlike in the bacterial growth case \cite{Judith} or viscous
fingering \cite{Mullins-Sekerka}.

From a broader perspective, we see this work as a first step towards a
systematic analysis of moving vortex fronts. The linear stability
analysis which we will develop can equally well be applied to
dynamical models which include coupling to the temperature or in which
the current-voltage characteristic is nonlinear.  For this
reason, we present the analysis in some detail for the relatively
simple case where the vortex velocity is linear with respect to the
magnetic field gradient and the current.  Even then, as we shall see,
the basic uniformly translating front solutions can still have
surprisingly complicated behavior. We find that the density of
vortices which penetrate the sample vanishes linearly for large enough
front velocities, but with a fractional exponent for front velocities
below some threshold velocity \cite{barenblatt}. Since the latter
regime appears to be physically less relevant, and since we do not
want to overburden the paper with mathematical technicalities, we will
focus our analysis on the first regime. As stated before, in this
regime we find that an anisotropy in the mobility without coupling to
the temperature does not give rise to an instability of the flux
fronts.

Our analysis will be aimed at performing the full stability analysis
of the fronts in the coupled continuum equations for the vortex
densities. Our procedure thus differs from the one of
\cite{Fisher,Fisher1} in which a sharp interface limit was
used.  In many physical systems it is often advantageous to map the
equations onto a moving boundary effective interface problem, in which
the width of the transition zone for the fields is neglected. One can
in principle derive the proper moving boundary approximation from the
continuum equations with the aid of singular perturbation theory. The
analogous case of the bacterial growth fronts \cite{Judith} indicates,
however, that such a derivation can be quite subtle for nonlinear
diffusion problems. Indeed it is not entirely clear whether the
assumptions used in the sharp interface limit of
Ref.~\cite{Fisher,Fisher1} are fully justified. For this reason, we
have developed an alternative and more rigorous stability analysis
which allows for a systematic study on fronts in vortex 
dynamics.

\subsection{The model}
The physical situation that we have in mind refers to a semi-infinite
2D thin film in which there is an initial uniform distribution of
vortices due to an external field {\bf H} applied along the $z$
direction.  By reversing and increasing the field, a front of vortices
of opposite sign penetrates from the edge of the film. We will refer
to the original vortices as antivortices with density $n^-$, and to
the ones penetrating in after the field reversal as vortices with
density $n^+$. In the region of coexistence of vortices and
antivortices, annihilation takes place. Vortices are driven into the
interior of the superconducting sample by a macroscopic supercurrent
{\bf J} along the $y$ direction due to the gradient in the density of
the internal magnetic field.  Flux lines then tend to move along the
direction $x$ transverse to the current under the influence of the
Lorentz force on each vortex (see e.g. \cite{giannirmp,gianni})
\begin{equation}
{\bf F}^\pm = \pm \frac{1}{c}{\bf J} \times {\bf \phi}_{0}\,{\bf e_z},
\label{lorentzforce}
\end{equation}
where ${\bf \phi}_{0}$ is the quantum of magnetic flux associated with
each Abrikosov vortex.  We consider the regime of pure flux flow in
which pinning can be neglected, while the viscous damping then gives
rise to a finite vortex mobility.  We follow a coarse-grained
hydrodynamic approach in which the fields vary on a scale much larger
than the distance between vortices.  Since the magnetic flux
penetrates in the form of quantized vortices, the total magnetic field
in the interior of the thin film can be expressed in a coarse graining
procedure through the difference in the density of vortices and
antivortices,
\begin{equation}
{\bf B} = (n^{+}-n^{-})\,\phi_0\,{\bf e_z}.
\label{field}
\end{equation} 
The dynamical equations for the fields of vortices and antivortices
are simply the continuity equations
\begin{eqnarray}
\frac{\partial n^+}{\partial t}&=&-{\bf \nabla} \cdot (n^{+}\,
{\bf{v}^+})-\frac{n^{+}n^{-}}{\tau},\nonumber\\
\frac{\partial n^-}{\partial t}&=&-{\bf \nabla} \cdot (n^{-}\,
{\bf{v}^-})-\frac{n^{+}n^{-}}{\tau},
\label{continuityeq}
\end{eqnarray}
where the second term on the right represents the annihilation between
vortices of opposite sign. Note that since vortices annihilate in
pairs, the total magnetic field $B_z$ is conserved in the annihilation
process. 
The annihilation terms depends on the recombination coefficient $\tau$;
a simple kinetic gas theory type estimate shows that $\tau^{-1}$ is of order 
of $v\xi_0$,
since the cross section of a vortex is of order $\xi_0$, the coherence 
length \cite{Shapiro}. 
The velocity $\bf{v}$ can be determined with the phenomenological formula
for the flux flow regime:
\begin{equation}
\eta {\bf{v}}^\pm=\pm\, {\bf{J}}\times\frac{\phi_0}{c}\,{\bf{e}_z},
\label{velox}
\end{equation}  
where the Hall term has been neglected with good
approximation for a case of a dirty superconductor \cite{gianni}. The
drag coefficient $\eta$ is given by the Bardeen-Stephen model
\cite{Bardeen} and generally depends on the temperature of the sample.
In this paper we neglect this important coupling to the temperature,
but we will allow the mobility (the inverse of the drag) to be
anisotropic.  In passing, we also note that the above linear relation
between the current {\bf J} and the flow velocity ${\bf v}^\pm$ is
often generalized to a nonlinear dependence \cite{Fisher1}. For
simplicity, we do not consider this case here, but our method can 
be extended to such situations.
 
For a type II superconducting material with a Ginzburg-Landau
parameter $\kappa \gg 1/ \sqrt{2}$, the magnetization of the sample
can be neglected, so that ${\bf B}\approx{\bf H}$.  Then, by using the
Maxwell equation (in which the term related to the displacement
currents has been neglected with good approximation)
\begin{equation}
{\bf J}= \frac{c}{4\pi}\,{{\bf\nabla}}\,\times\,{\bf B} ,
\end{equation}
together with (\ref{field}) and (\ref{velox}), and substituting into
(\ref{continuityeq}), we get:
\begin{eqnarray}
\frac{\partial n^+}{\partial t}&=&D\,{\bf \nabla} \cdot 
(n^{+}\,{\bf \nabla}(n^{+}-n^{-}))-\frac{n^{+}n^{-}}{\tau},\label{dynamiceq1}\\
\frac{\partial n^-}{\partial t}&=&D\,{\bf \nabla} \cdot
(n^{-}\,{\bf \nabla}(n^{-}-n^{+}))-\frac{n^{+}n^{-}}{\tau}.
\label{dynamiceq2}
\end{eqnarray}
where the coefficient D is given by ${\ds D=\phi_0^2 /(4 \pi \eta)}$.
This is the system of non-linear differential equations which governs
the dynamics of the vortex-antivortex front. 
The situation that we will study in our analysis is the following.
We consider a front of vortices which propagates into the
superconducting thin film from the left edge at $x=-L_x$ in the
positive $x$ direction. At $x=-L_x$, we impose the boundary condition
that the density of vortices $n^+$ is ramped up linearly in time,
$n^+(-L_x,t) = R t$. This corresponds to the field going up linearly,
 just as in the Bean critical state \cite{bean}.
We impose also that far right at $x\to\infty$, $n^+$
vanishes while $n^-$ approaches a constant value $n_\infty$.
Through a rescaling of time and length variables, the coefficients of the
equations (\ref{dynamiceq1}) and (\ref{dynamiceq2}) can be set to
unity. 
In particular, it is convenient to rescale the time
and length variables according to the following transformation:
\begin{eqnarray}
t &\rightarrow& \frac{t \,n_\infty}{\tau},\nonumber\\
x&\rightarrow&\frac{x}{l_0}= x \sqrt{\frac{4\pi\eta}{\phi_0^2\tau}},\nonumber\\
n&\rightarrow&\frac{n}{n_{\infty}}.\label{scale}
\end{eqnarray}
We will henceforth analyze the equations (\ref{dynamiceq1}) 
and (\ref{dynamiceq2}) with $D=1 $ and $\tau=1$.\\
As we already mentioned in the introduction, and as we shall see in detail 
below, the above continuum equations have a mathematical singularity at the 
point where $n^+$ vanishes. Of course, in reality there cannot be such a true 
singularity and our continuum coarse-grained model breaks down at scales 
of the order of the London penetration depth.
In particular, the derivative of the magnetic field 
and thus the current $\bf{J}$ are not discontinuous 
with respect to the space variable, but they decrease exponentially 
in a distance approximately equal to the penetration depth.
Effects like thermal diffusion, the finite core size, and the 
nonlocal relations which are neglected in the London approximation
 all play a role there, and the Ginzburg-Landau 
equation would provide a more appropriate starting point. Clearly, if the 
dynamical behavior of our continuum model would be very sensitively dependent 
on the nature of the singularity, then this would be a sign that the physics 
at this cutoff scale would really strongly affect the dynamically
relevant long-wavelength dynamics. 
In practice, however, this is not the case. First of all, our 
method to do the linear stability analysis is precisely aimed at making sure 
that the singularities at the level of the continuum equations does not mix 
with the behavior or pertubations of the front region. 
Secondly, as we shall see there are no instabilities on scales of the order 
of the microscopic cutoff provided by the London penetration depth.

\subsection{The Method}
In our analysis, we first study a planar front which propagates with a
steady velocity \emph{v} along the \emph{x} direction. By considering
the propagation of the front in the comoving frame, we get a system
of ordinary differential equations (ODEs) for the vortex and antivortex density fields.  The
derivation of the uniformly translating solution is discussed in
Section~\ref{due}.  As we will see, the profile that corresponds to
the planar front for the density of vortices is singular. In
particular, in the regime on which we will focus, the derivative of
the vortex density is discontinuous at the point where the field
vanishes, while in the low-velocity regime there are higher order
singularities.  As a consequence of this nonanalytic behavior, the
numerical integration of the equations has to be done with care near
the singular point.

In Section~\ref{tre}, we perform a linear stability analysis of the
planar solution. A proper ansatz consists here of two contributions: a
perturbation in the line of the singular front and a perturbation of
the density field.  As we will see, the presence of an in-plane
anisotropy means that the (anti)vortex flow velocity is no longer in
the same direction as the driving force acting on the
(anti)vortices. Hence, contrary to the isotropic case, we have to
consider a component of the velocity perpendicular to the driving
force. The viscosity is thus represented by a non-diagonal tensor and
depends on the angle between the direction of propagation of the front
and the fast growth direction given by the anisotropy.  By applying a
linear stability analysis we get a system of equations for the fields
representing the perturbation.  Through a shooting method, and by
matching the proper boundary conditions, we are then able to determine
a unique dispersion relation for the growth rate of the perturbation.
In Section~\ref{quattro} we treat the case of a stationary front,
with a velocity $v=0$. Contrary to the case of a moving front, no
singularity in the profiles of the fields is present and the analysis
can be carried out in the standard way.  }

\section{The planar front}

{
\subsection{The equations and boundary conditions}
\label{due}
\begin{figure}[tb]
\includegraphics[width=7.5 cm]{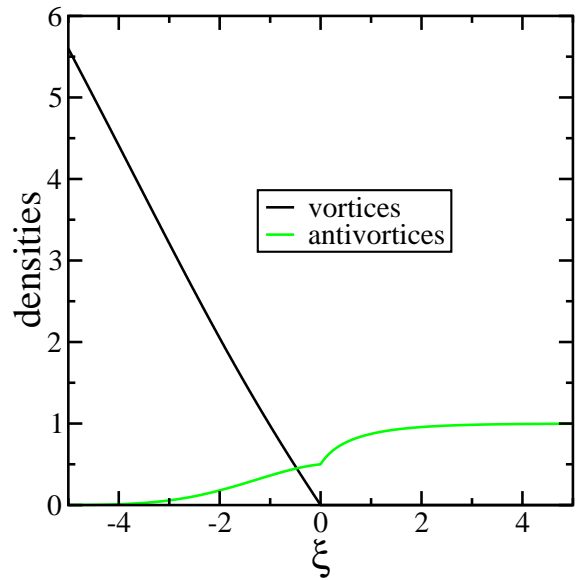}
\caption{Profile of the planar front for the density of the
 vortices ($n^+$) and antivortices ($n^-$) for the case $v=1$.}
\label{fig.planarfront}
\end{figure}

In this section, we analyze the planar uniformly translating front
solutions $n^+=n^+_0(x-vt)$, $n^-=n^-_0(x-vt)$ which are the starting
point for the linear stability analysis in the next section.  We refer
to the system in a comoving frame in which the new coordinate is
traveling with the velocity $v$ of the front, ${\ds \xi=x-vt}$.  The
temporal derivative then transforms into ${\ds
\partial_t|_x=\partial_t|_\xi-v\partial_\xi}$. Since the front is
uniformly translating with velocity $v$, the explicit time
derivative vanishes. In the comoving frame system, 
we consider $\xi$ to vary in the spatial interval $[-L, +\infty]$.
The equations (\ref{dynamiceq1}) and
(\ref{dynamiceq2}) become:
\begin{eqnarray}
-v\frac{dn_0^+}{d\xi}&=&\frac{d}{d\xi}n_0^{+}\,
\frac{d}{d\xi}(n_0^{+}-n_0^{-})-n_0^{+}n_0^{-},\label{comovingframe1}\\
-v\frac{dn_0^-}{d\xi}&=&\frac{d}{d\xi}n_0^{-}\,
\frac{d}{d\xi}(n_0^{-}-n_0^{+})-n_0^{+}n_0^{-}.\label{comovingframe2}
\end{eqnarray}
This is a system of two ODEs of second order.  Motivated by the
physical problem we wish to analyze, the relevant uniformly
translating front solutions obey the following
boundary conditions at infinity:
\begin{equation}
\begin{array}{ll}
\ds\lim_{\xi \rightarrow +\infty} n_0^- = n_{\infty},\hspace{.7 cm} 
&{\ds \lim_{\xi \rightarrow +\infty} \frac{dn_0^-}{d\xi}=0},\\
\ds\lim_{\xi \rightarrow +\infty} n_0^+ = 0,\hspace{1.1 cm} 
&{\ds \lim_{\xi \rightarrow +\infty} \frac{dn_0^+}{d\xi}=0}.
\label{bcinfty+}
\end{array}\end{equation}
It is important to note that the constant $n_\infty$ can actually be
set to unity: by rescaling the density fields as well as space and
time, any problem with arbitrary $n_\infty$ can be transformed into a
rescaled problem with $n_\infty=1$. The stability of fronts therefore
does not depend on $n_\infty$, and in presenting numerical results we
always use the freedom to set $n_\infty=1$.

On the left, the density of vortices $n^+$
increases linearly with time with sweeping rate $R$.  After a
transient time, because of the annihilation process, the field $
n_0^-$ and its derivative vanish.  The dynamical equation
(\ref{comovingframe1}) for the $n^+$ field then yields
\begin{equation}
\frac{dn_0^+}{d\xi}=-v + \mathcal{O}\left(\frac{1}{n_0^+}\right),
\label{asymptn+}
\end{equation}
i.e., we recover the well known critical state result \cite{bean} that
in the absence of antivortices the penetrating $n^+$ field varies
linearly with slope $-v$. Requiring that this matches the
boundary condition $n^+(-L_x,t)=Rt$ for large times at $\xi=-L$
 then immediately yields that $R = v^2$. 
It can be easily derived that the density of antivortices decays with a 
Gaussian behavior on the left.
By using indeed the relation (\ref{asymptn+}) for large distances 
 and substituting it in (\ref{comovingframe2}), we get:
\begin{equation}
n_0^-\approx Ae^{-\xi^2/4}\label{gaussiann-}.
\end{equation}
Since the
analysis of the planar front profiles and of their stability is
naturally done in the comoving $\xi$ frame, we will in practice use a
semi-infinite system in the $\xi$ frame, and impose as boundary
conditions at $\xi=-L$
 \begin{equation}
\begin{array} {ll}
\ds\lim_{\xi \rightarrow -L}{\ds n_0^-=0},\hspace{.3 cm}  
&\ds\lim_{\xi \rightarrow -L}{\ds \frac{dn_0^-}{d\xi}=0},\\
\ds\lim_{\xi \rightarrow -L}{\ds n_0^+= const \gg 1},\hspace{0.5 cm}  
&\ds\lim_{\xi \rightarrow -L}{\ds \frac{dn_0^+}{d\xi}}=-v  .
\label{bc-L}
\end{array}
\end{equation}
Of course, in any calculation we have to make sure that $L$ is taken
large enough that the profiles $n_0^\pm$ have converged to their
asymptotic shapes.

\subsection{Singular behavior of the fronts}
\label{sbotf}
Effectively, Eqs.~(\ref{dynamiceq1}-\ref{dynamiceq2}) and
(\ref{comovingframe1}-\ref{comovingframe2}) have the form of diffusion
equations whose diffusion coefficient vanishes linearly in the
densities $n^+$ and $n^-$.  As already mentioned before, it is well
known, from e.g. the porous medium equation
\cite{porous1,porous2,porous3}, that such behavior induces singular
behavior at the point where a density field vanishes (see
e.g. Ref.~\cite{Bender}). Because we are looking at fronts moving into
the region where $n^+=0$, in our case the singularity is at the point
where the $n^+$ density vanishes.  Let us choose this point as the
origin $\xi=0$. Then the relevant front solutions have $n^+(\xi)=0 $
for all $\xi>0$; see Fig.~\ref{fig.planarfront} \cite{note3}.

Because $n^-_0(0)\neq 0$, the prefactor of the highest derivative in
the $n^-$ equation does not vanish at $\xi=0$, and hence one might
naively think that $n^-$ is nonsingular at this point. However,
because of the coupling through the diffusion terms, this is not
so. By integrating Eq.~(\ref{comovingframe2}) over an interval
centered around $\xi=0$ and using that the field values $n^+_0$ and
$n^-_0$ are continuous, one immediately obtains that
\begin{equation}
\lim_{\Delta \xi\rightarrow
0}\left.\left(\frac{dn_0^+}{d\xi}-\frac{dn_0^-}{d\xi}\right)\right|_{-\Delta
\xi}^{\Delta \xi}=0. \label{jump}
\end{equation}
Physically, this constraint expresses the continuity of the derivative
of the coarse-grained magnetic field (\ref{field}). Mathematically, it
shows that any singularity in $n^+_0$ induces precisely the same
singularity in $n^-_0$: to lowest order the two singularities
cancel. Fig.~\ref{fig.planarfront} illustrates this: one can clearly
discern a jump in the derivative of $n^-_0$ at the point where $n^+_0$
vanishes with finite slope.

Before we analyze the nature of the singularity in more detail, we
note that because of the nonanalytic behavior at $\xi=0$, it is
necessary to analyze the region $\xi<0$ where $n^+_0\neq 0$ separately
from the one at $\xi>0$ where $n^+_0=0$. In the latter regions, the
equations simplify enormously, as the remaining terms in
Eq.~(\ref{comovingframe2}) can be integrated immediately. Upon
imposing the boundary conditions (\ref{bcinfty+}) at infinity, this
yields
\begin{equation}
\frac{dn_0^-}{d\xi}=-v\frac{(n_0^--n_{\infty})}{n_0^-},
\hspace*{0.5cm} \xi>0.\label{positivexi}
\end{equation}

Let us now analyze of the nature of the singularity at $\xi=0$.  As
the effective diffusion coefficient of the $n^+$-equation is linear in
$n^+$, analogous situations in the porous medium equation suggest that
the field $n^+$ vanishes linearly.  This motivates us to write for
$-1\ll \xi<0$ \cite{note4}
\begin{equation}
\begin{array}{l}
{\ds n^+_0(\xi)  =  A^+_1 \xi  + A^+_2 \xi^{2} +\cdots ,}\\
{\ds n^-_0(\xi)   =  A^-_1 \xi   + A^-_2  \xi^{2} +\cdots +
n^-_{\rm an}(\xi) ,} 
\end{array} \label{expansion1}
\end{equation}
where $n^-_{\rm an}(\xi) $ is the analytic function which obeys
Eq.~(\ref{positivexi}) for {\em all} $\xi$. Clearly, the continuity
condition (\ref{jump}) immediately implies
\begin{equation}
A^+_1=A^-_1  . \label{Aeqs}
\end{equation}
If we now substitute the expansion (\ref{expansion1}) with
(\ref{Aeqs}) into Eq.~(\ref{comovingframe1}) for $n^+_0$ we get by
comparing terms of the same order:
\begin{equation}
\begin{array}{l}
{\ds {\mathcal O}(1):~~~   A^+_1 (v  - n^{-\prime}_{\rm an} )  =0,  }\\
{\ds  {\mathcal O} (\xi) : ~~~ 4 (A^+_2 - A^-_2) - 
2 n^{-\prime\prime}_{\rm an} - n^-_{\rm an}=0. } 
\end{array} \label{expansion2}
\end{equation}
Here $n^{-\prime}_{\rm an}= dn^-_{\rm an}/d\xi|_{\xi=0}$,
etc. Likewise, if we substitute the expansion into
Eq.~(\ref{comovingframe2}) for $n^-_0$, we get
\begin{equation}
{\ds {\mathcal O}(1):~~~   2v A^-_1 - 2n^{-}_{\rm an}(A^+_2-A^-_2 )  =0, }
 \label{expansion3}
\end{equation}
since the term of order unity involving $n^-_{\rm an}$ cancels in
view of (\ref{positivexi}).  Higher order terms in the expansion
determine the coefficients $A_2^+ $ and $A_2^-$, and other terms like
$A_3^\pm$ separately, but are not needed here.  Together with
(\ref{positivexi}), the above equations
(\ref{expansion2}-\ref{expansion3}) immediately yield
\begin{equation}
\begin{array}{rl}
{\ds   n_{\rm an}^{-\prime} } = & {\ds v,}\\
{\ds n_{\rm an}^-(0)}= & {\ds 1/2} , \\

{ \ds  A^+_1 = A^-_1}  = & {\ds -v +\frac{1}{16v} , }
\end{array}\label{bc0}
\end{equation}
where for convenience we have now put $n_\infty=1$. 

There are two curious features to note about the above result. First
of all, $n^+_0$ always vanishes at the point where $n^-_0$ is half of
the asymptotic value $n_\infty$ at infinity. Secondly, note that
$A^+_1$ is negative for $v \geq 1/4$ and positive for $v<1/4$. Since the
vortex density $n^+$ has to be positive, we see that these uniformly
translating front solutions can only be physically relevant for $v \geq
1/4$!

Since the front velocity in this problem is not dynamically selected
but {\em imposed} by the ramping rate $R=v^2$ at the boundary, we do
expect physically realistic solutions with $v<1/4 $ to exist. In fact,
it does turn out that in this regime the nature of the singularity
changes: instead of vanishing linearly, $n^+_0$ vanishes with a
$v$-dependent exponent. Indeed, if we write for $-1\ll \xi<0$
\cite{note4}
\begin{eqnarray}
n^+_0(\xi) & = &  |\xi|^{\alpha}(A^+_1 + A^+_2 \xi +\cdots ) ,\\
n^-_0(\xi)  & = &  |\xi|^{\alpha} (A^-_1 + A^-_2 \xi +\cdots ) +
n^-_{\rm an}(\xi) ,
\end{eqnarray}
and substitute this into the equations, then, in analogy with the result
above, we find
\begin{equation}
\begin{array}{rl}
{ \ds n_{\rm an}^{-\prime}}   = & {\ds v,}\\ 
{ \ds n_{\rm an}^-(0)}= & {\ds 1/2} , \\
{ \ds  A^+_1} =& {\ds  A^-_1}, \\
{\ds \alpha} = & {\ds \frac{1}{8v^2}-1>1 }, ~~~(v<1/4),
\end{array}\label{bc2}
\end{equation}
while again for $\xi>0$ $n^+_0$ vanishes. A singular behavior with
exponent depending on the front velocity $v$ is actually quite
surprising for such an equation \cite{barenblatt}. However, one should
keep in mind that this behavior is intimately connected with the
initial condition for the $n^-$ vortices. If one starts with a case
where $n^-$ does not approach a constant asymptotic limit on the far
right, but instead increases indefinitely, one will obtain solutions
where $n^+$ vanishes linearly. For this reason, and in order not to
overburden the analysis with mathematical technicalities, from
here on we will concentrate the analysis on the regime $v \geq 1/4$.

Since our study will limit the stability analysis to fronts with velocity 
$v \geq 1/4$ in our
dimensionless variables, let us check how the scale 
that we consider relates with the realistic values of flux flow velocities.
By considering relations (\ref{scale}) the velocities are measured in units of:
\begin{eqnarray}
v_0= \frac{l_0 n_{\infty}}{\tau}&=& c\sqrt{\frac{H_{\infty}}{H_{c2}}}\sqrt{\frac{\rho_n n_{\infty}v\xi_0}{4\pi}},\nonumber\\
&\approx&c\sqrt{\frac{\xi_0^2}{a^2}}\sqrt{\frac{\rho_n v\xi_0}{4\pi a^2}},\label{velocity1}
\end{eqnarray}
where we have expressed the viscosity $\eta$ in terms of the upper 
critical field $H_{c2}$ and the normal state resistivity $\rho_n$
by using \cite{Bardeen}.
Furthermore, $a$ is the distance between vortices for $\xi\rightarrow\infty$;
thus, since  $H_{\infty}= \phi_0 n_{\infty}\approx  \phi_0/a^2$, 
it follows that $H_{\infty}/H_{c2}\approx\xi_0^2/a^2$.
For the constant $\tau$ we have used the estimate $\tau^{-1}=v\xi_0$ 
discussed after Eq. (\ref{continuityeq}).
We can then rewrite (\ref{velocity1}) as 
\begin{equation}
v\approx c^2 \frac{\xi_0^3}{4\pi a^4}\rho_n.
\end{equation}
By considering typical values in Gaussian units
$\rho_n\approx~10^{-16} s$ for 
the resistivity of the material, a coherence length 
$\xi_0\approx 2\,10^{-7}$ cm for high-$T_c$ compounds
and a magnetic field $H_{\infty}\approx 20 G$ 
(which corresponds to a length $a\approx 10^{-4}$ cm), our velocity is then
 measured in units of $v\approx 1\,cm/s$. 
This velocity scale is much less than values found
typically in the flux flow regime, since in the presence of instabilities
fronts of vortices can propagate with much higher velocities of 
order $10^4-10^6 cm/s$ \cite{Shantsev2}. Thus, the regime $v \geq 1/4$ 
is indeed the physical relevant one.
   
\subsection{Sum and difference variables}
At first glance, the equations look like two coupled second order
equations. However, there is more underlying structure due to the fact
that the annihilation term does not effect the difference $n^+-n^-$.
In order to integrate the set of equations
(\ref{comovingframe1}-\ref{comovingframe2}), it is convenient to
consider the following transformations in the variables related to the
sum and difference of the density fields:
\begin{equation}
\begin{array}{ll}
{\ds D}&={\ds n^+-n^- ,}\\
{\ds S}&={\ds n^++n^-. }
\end{array}
\label{SDtransf} 
\end{equation}
In these variables, the equations become
\begin{eqnarray}
-v\frac{dD_0}{d\xi}&=&\frac{d}{d\xi}S_0\frac{dD_0}{d\xi}\label{SD1},\\
-v\frac{dS_0}{d\xi}&=&\frac{d}{d\xi}D_0\frac{dD_0}{d\xi} 
-\frac{S_0^2-D_0^2}{2}.
\label{SD2}
\end{eqnarray} 
By numerically integrating (\ref{SD1}-\ref{SD2}) and looking for the
solutions which satisfy the boundary conditions above, we obtained the
uniformly translating front solutions.  As Fig.~\ref{fig.planarfront}
illustrates for $v=1$, the profile is singular at the point where
the density of the $n^+$ field vanishes linearly, in agreement with
the earlier analysis.

Because of this singularity, the numerical integration of the set
(\ref{SD1}-\ref{SD2}) is quite nontrivial.  In particular, because of
the discontinuity in the derivative of the $n^+$ field, the system
(\ref{SD1}-\ref{SD2}) effectively needs to be solved only in the
interval $[-L, 0[$, as the matching to the behavior for $\xi>0$ has
already been translated into the boundary conditions (\ref{bc0}).  The
first equation can be straightforwardly integrated and by combining it
with the second, the set reduces to
\begin{equation}
\begin{array}{ll}
{\ds \frac{dD_0}{d\xi}} &={\ds \frac{-v(D_0+n_{\infty})}{S_0}},\\
{\ds \frac{dS_0}{d\xi} } &= {\ds \frac{{ S_0(v(2D_0+n_{\infty})){\ds
\frac{dD_0}{d\xi}}+(S_0^4-S_0^2D_0^2)/2}}{{\ds vS_0^2+
(v(D_0^2+n_{\infty}D_0))}}}.
\end{array}
\label{SD3}
\end{equation}
One can easily verify that in this formulation, the expression on the
right hand side is indefinite at the singular point $\xi=0$, as both
the terms in the numerator and denominator vanish.  In order to
evaluate the expression, it is then necessary to perform an expansion
of the numerator and denominator around the critical point values
$S=-D=n_\infty/2$. From such an analysis one can then recover the
relations (\ref{bc0}) which we previously obtained from a
straightforward expansion of the original equations. Numerically, we
integrate the equations by starting slightly away from the singular
point with the help of the results from the analytic expansion.

\section{Front propagation in the presence of anisotropy}
{\label{tre}
\subsection{Dynamical equations}

As mentioned before, we are interested in the effect that an
anisotropy in the vortex mobility could have on the stability of the
front.  In particular, the motivation for such an investigation is the
experimental evidence that an instability for a flux-antiflux front
was found in materials with an in-plane $ab$ anisotropy, such as for
example $YBa_2Cu_3O_{7-\delta}$ \cite{Koblischka}.

In a material characterised by an in-plane anisotropy, the effective
viscous drag coefficient depends on the direction of propagation of
the front. More precisely, the mobility defined in (\ref{velox}) then
becomes a non-diagonal tensor.  This leads to a non-zero component of
the velocity $v$ perpendicular to the driving Lorentz force. We want
to investigate whether the non-collinearity between the velocity and
the force is responsible for an instability of the flux-antiflux
interface.  In the presence of anisotropy, the phenomenological
formula (\ref{velox}) then has to be replaced by:
\begin{equation}
{\bf{v}}=\hat{\eta}^{-1}{\bf{F}}=\Gamma R^{-1}\left(\begin{array}{cc}1&0\\
0&\alpha\end{array}\right)R{\bf{F}},
\end{equation}
where $\Gamma$ is a constant, $\alpha$ represents the anisotropy
coefficient and $R$ is the rotation matrix corresponding to an angle
$\theta$ between the direction of propagation of the front $x$ and the
principal axes $x'$ of the sample.  The coefficient $\alpha$ varies in
the range [0,1] with the limiting case of infinite anisotropy
corresponding to $\alpha \to 0$. For $\alpha=1$ the isotropic case is
recovered.  The matrix $\hat{\eta}^{-1}$ is given in particular by
\begin{equation} 
\hat{\eta}^{-1}=\Gamma\left(\begin{array}{cc}\cos^2{\theta} + 
\alpha \sin^2{\theta}&\cos{\theta} \sin{\theta}(1- \alpha)\\
\cos{\theta} \sin{\theta}(1- \alpha)&\ds\alpha\cos^2{\theta}+
\sin^2{\theta}\end{array}\right),
\end{equation}
The dynamical equations for the fields $n^+$ and $n^-$ in the presence
of anisotropy generalize to
\begin{eqnarray}
\frac{\partial n^\pm }{\partial t}&=&\frac{\partial }{\partial
x}\left(n^{\pm }\frac{\partial }{\partial
x}\left(n^{\pm }-n^{\mp }\right)\right)\nonumber 
\\ & + & p\frac{\partial }{\partial
y}\left(n^{\pm }\frac{\partial }{\partial y}
\left(n^{\pm }-n^{\mp}\right)\right)\nonumber\\
               &+&k\frac{\partial }{\partial x}\left(n^\pm \frac{\partial
}{\partial y}\left(n^\pm -n^\mp \right)\right)\nonumber\\
               &+&k\frac{\partial }{\partial y}\left(n^\pm
\frac{\partial }{\partial x}\left(n^\pm -n^\mp \right)\right)
                 -n^{+}n^{-}\label{aniso1},
\end{eqnarray}
where the length and time variables have been rescaled and the
elements $k$ and $p$ depend on the angle $\theta$ through the
formulas:
\begin{eqnarray}
k= {\ds\frac{ \cos{\theta} \sin{\theta}(1- \alpha)}
{ \cos^2{\theta} + \alpha \sin^2{\theta}}},\mbox{\hspace{.5 cm}}
p= {\ds\frac{ \alpha\cos^2{\theta}+\sin^2{\theta}}
{ \cos^2{\theta} + \alpha \sin^2{\theta}}}\label{matrixelements}.
\end{eqnarray}
Starting from an initially planar profile derived in
Section~\ref{due}, we want to study the linear stability of the front
of vortices and antivortices by performing an explicit linear
stability analysis on Eq.~(\ref{aniso1}).

\subsection{The linear stability analysis}
\begin{figure}[b]
\includegraphics[width= 6 cm]{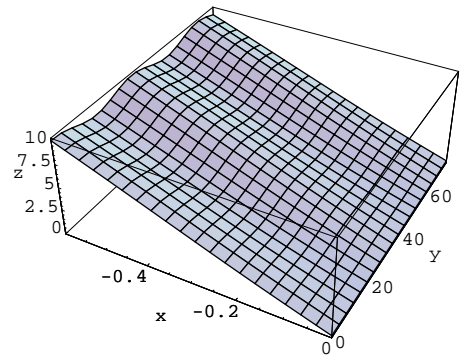}
\includegraphics[width= 6 cm]{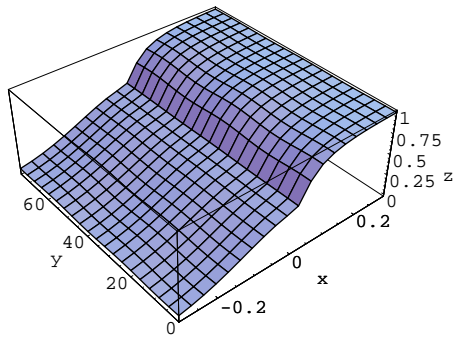}
\caption{Perturbed front profile for the vortex and antivortex
density field.  The fronts propagate in the $x$ direction and has a
sinusoidal modulation in the $y$ direction}
\label{perturbation.fig}
\end{figure}

As we have already mentioned in earlier sections, our linear
stability analysis differs from the standard one, due to the presence
of a singularity.  The type of perturbation that we want to consider
should not only involve the profile in the region where $n^+$
vanishes, but should also in particular involve the geometry of the
front. In other words, as Fig.~\ref{perturbation.fig} illustrates, we
want to perturb also the location of the singular line at which the
density $n^+$ vanishes. As discussed in more detail in \cite{Judith},
the proper way to implement this idea is to introduce a modulated
variable
\begin{equation}
\zeta(\xi,y,t)= \xi+\epsilon e^{iqy + \omega t +i\Omega t }
\label{shiftxi}
\end{equation}
and then to write the densities in terms of this ``comoving''
modulated variable. Of course, the proper coordinate is the real
variable $\mbox{Re}\,\zeta$.  However, when we expand the functions in
Fourier modes and linearize the dynamical equations in the amplitude
$\varepsilon$, each Fourier mode can be treated separately.  Thus, we
can focus on the single mode with wavenumber $q$ and amplitude
$\epsilon$ and then take the real part at the end of the calculation.
The profiles of the fields $n^+$ and $n^-$ are now perturbed by
writing
\begin{eqnarray}
n^+(\zeta,y,t)&=n^+_0(\zeta)+\epsilon (n^+_1 + i n^+_2)(\zeta) 
e^{iqy+\omega t+i\Omega t}&,\label{+perturbation}\\
n^-(\zeta,y,t)&=n^-_0(\zeta)+\epsilon (n^-_1+i n^-_2)(\zeta)) 
e^{iqy+\omega t+i\Omega t}&,\label{-perturbation}
\end{eqnarray}     
where $n^{+}_0$ and $n^-_0$ are simply the planar front profiles
determined before. Note that since we write these solutions as a
function of the modulated variable $\zeta$, even the first term already
implies a modulation of the singular line. Indeed, the standard
perturbation ansatz would fail for our problem because of the singular
behavior of the front. The usual ansatz of a stability calculation
 \begin{equation}
n^+(\xi,y,t)= n^+_0(\xi)+
\epsilon (n^+_1+ in^+_2)(\xi) e^{iqy+ \omega t +i\Omega t}
\end{equation}
 only works if the unperturbed profiles are smooth enough and not
vanishing in a semi-infinite region. If we impose on our corrected
linear stability analysis the conditions
\begin{equation}
\frac{n^{+}_1+ in^{+}_2}{n^{+}_0}\hspace{.3 cm}
\mbox{ bounded}\hspace{.3 cm} \mbox{and}\hspace{.3 cm}
\frac{n^{-}_1+i n^{-}_2}{n^{-}_0} \hspace{.3 cm}\mbox{ bounded}
\label{bounds}
\end{equation}
then as $\varepsilon\to 0$ the perturbations can be considered small
everywhere, plus we allow for a modulation of the singular line
\cite{Judith}.

 We next linearize the equations (\ref{aniso1}) around the uniformly
 translating solution according to
 (\ref{+perturbation}-\ref{-perturbation}).  We obtain a set of 4
 linearized ODEs for the variables ${D_1,D_2,S_1,S_2}$, which
 correspond, respectively, to the real and imaginary parts of the
 difference and sum variables introduced in (\ref{SDtransf}). These
 equations, which are reported in the Appendix, depend also on the
 unperturbed profiles $D_0, S_0$, which are known from the derivation
 in Section~\ref{due}.  Moreover, there is an explicit dependence on
 the parameters $q, \omega, \Omega$.

In order to analyze the stability of the front of vortices and
antivortices, the dispersion relation $\omega (q), \Omega(q)$ must be
derived.  This can be determined with a shooting method: for every
wavenumber $q$ there is a unique value of the growth rate $\omega$ and
frequency $\Omega$ which satisfies the boundary conditions related to
the perturbed front. If the growth rate is positive, a small
perturbation will grow in time, thus leading to an instability.

\subsection{The shooting method}
{
The singularity of the front makes the numerical integration difficult
to handle, as in the case of the planar front.  In view of the
relations (\ref{bounds}), the boundary conditions
\begin{equation}
n^+_1=0, \mbox{\hspace{1 cm}}n^+_2=0,\nonumber
\end{equation}
have to be imposed for $\zeta = 0$.  These yield the boundary
conditions for the variables $D_1,S_1,D_2,S_2$
\begin{equation}
S_1=-D_1, \mbox{\hspace{1 cm}}S_2=-D_2.\label{D1-S1}
\end{equation}
Moreover, by substituting these boundary conditions and the relations
 (\ref{bc0}) for the unperturbed fields in the linearized equations
 for ${D_1,D_2,S_1,S_2}$, the following relations can be derived for
 $\zeta$ vanishing from the left \cite{note5}:
\begin{eqnarray}
\left.  \frac{dD_1}{d\zeta}\right|_{0^-}& = &\omega +qkD_2(0),\label{derD1}\\
  \left. \frac{dD_2}{d\zeta}\right|_{0^-}& = & \Omega-qkD_1(0)
-\left. 2qk\frac{dD_0}{d\zeta}\right|_{0^-}.\label{derD2}
\end{eqnarray}
An explicit expression for the derivative of the sum of the real and
imaginary part of the perturbations $S_1, S_2$ can also be derived
from the equations reported in the Appendix.  In particular, these have
the following generic form
\begin{equation}
\frac{dS_1(\zeta)}{d\zeta} = \frac{{\cal N}_1(\zeta)}{{\cal
D}_1(\zeta)},\mbox{\hspace{1.0 cm}} \frac{dS_2}{d\zeta} 
= \frac{{\cal N}_2(\zeta)}{{\cal D}_2(\zeta)},\label{Sder}
\end{equation}
which is similar in structure to Eqs.~(\ref{SD3}): ${\cal N}_1,{\cal
 D}_1,{\cal N}_2,{\cal D}_2$ depend on $\zeta$ through the set of
 functions
\[\left(D_0,S_0,\frac{dD_0}{d\zeta},\frac{dS_0}{d\zeta},
D_1,S_1,\frac{dD_1}{d\zeta},D_2,S_2,\frac{dD_2}{d\zeta}\right),
\label{functions}\]
and on the parameters $q,\omega,\Omega$. 

The equations (\ref{Sder}) are not defined at the singular point. By
substituting the boundary conditions given by
(\ref{D1-S1}-\ref{derD2}}), both the numerators ${\cal N}_1,{\cal
N}_2$ and the denominators ${\cal D}_1,{\cal D}_2$ vanish. Again, as
with (\ref{SD3}), we encounter the problem of dealing with the
singularity at $\zeta =0$.  This difficulty can be overcome in the
same way as in Section~\ref{sbotf} for the derivation of the planar
front profile.
In particular,  we can not start the integration at the singular 
point, but we have to start the backwards integration at some small  
distance on the left of $\zeta= 0$.
We do so by first obtaining the derivatives of the fields $S_1$ and $S_2$
 analytically through the expansion of the equations (\ref{Sder})
around the critical point. 
In the limit $\zeta \rightarrow 0$, this yields the 
following self-consistency condition for the derivatives.
\begin{equation}
\left. \frac{dS_1}{d\zeta}\right|_{0^-}= \frac{{\cal
N}'_1|_{0^-}}{{\cal D}'_1|_{0^-}},\mbox{\hspace{0.5
cm}}\left. \frac{dS_2}{d\zeta}\right|_{0^-}= \frac{{\cal
N}'_2|_{0^-}}{{\cal D}'_2|_{0^-}}, \label{Sder1}
\end{equation}
where ${\cal N}'_1,{\cal N}'_2,{\cal D}'_1,{\cal D}'_2$ denote the
derivatives of the corresponding functions evaluated at the singular
point.
Once these are solved and used in the numerics, the integration can be 
carried out smoothly.
\begin{figure}[t]
\includegraphics[width= 7.0 cm]{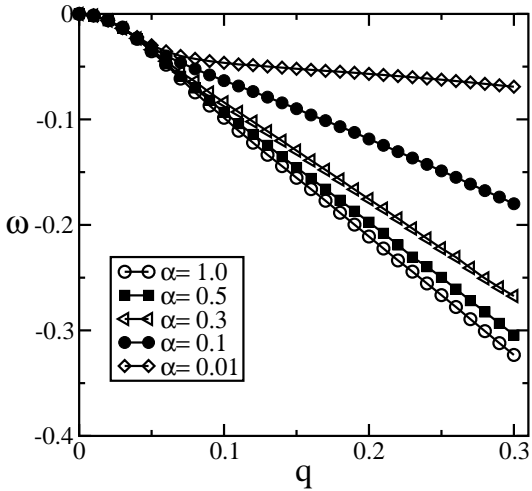}
\caption{Dispersion relation $\omega(q)$ for different values of 
anisotropy 
coefficient $\alpha$ and a velocity $v=1.0$.}
\label{fig.aniso}
\end{figure} 
Because of the singularity at the point $\zeta=0$, the derivative of
the perturbed fields are not continuous there and a relationship for
the discontinuity in the derivatives can be derived as was the case
for the unperturbed fields. In particular, the expression (\ref{jump})
is generalized for the perturbed field.  This implies that the
derivative of the total magnetic field is again continuous even at the
singularity.

From the equations for the perturbed fields given in the
Appendix, the boundary conditions at $\xi=-L$
can be derived. Just like  the unperturbed field for the antivortex
density vanishes on the left with a Gaussian behaviour 
 according to (\ref{gaussiann-}), also the perturbations $n_1^-$ and
$n_2^-$ vanish as a Gaussian, i.e. faster then an exponential. 
 
Moreover, since the density
of vortices increases linearly asymptotically,
we can retain in the equations only terms which are proportional to
the density of vortices $n_0^+$.  From this we get the following
equation for the density of the perturbation $\delta n^+= n_1^+ + i
n_2^+$ for $\zeta \ll -1$
\begin{equation}
\frac{d^2\delta n^+}{d^2 \zeta}+2i 
qk\frac{d\delta n^+}{d \zeta}-pq^2\delta n^+=pq^2\frac{d n_0^+}{d\zeta}.
\label{asymptequation}
\end{equation}  
The solutions of this equation which do not diverge are of the form
\begin{equation}
\delta n^+ = - \frac{dn_0^+}{d \zeta}+C e^{\lambda \zeta}, \hspace{.2
cm}\lambda = i qk +\sqrt{(q^2(p-k^2))},\label{asymptsol}
\end{equation} 
where $C$ is an arbitrary constant and $k$ and $p$ represent the
coefficients of anisotropy defined in (\ref{matrixelements}). Thus,
the perturbations decay on the left of the film with a decay length
$\zeta_0$, such that
\begin{equation}
\frac{1}{\zeta_0}= q \sqrt{p-k^2}.\label{declength}
\end{equation} 
Note that the decay length becomes very large for small $q$; this
type of behavior is of course found generically in diffusion limited
growth models. Technically it means that we need to be careful to take
large enough systems to study the small-$q$ behavior. From the numerical
integration it was verified that Eqs.
(\ref{asymptsol}) and  (\ref{declength}) describe correctly the behavior 
of $\delta n^+$ at large distance. 
\begin{figure}[t]
\includegraphics[width= 7.0 cm]{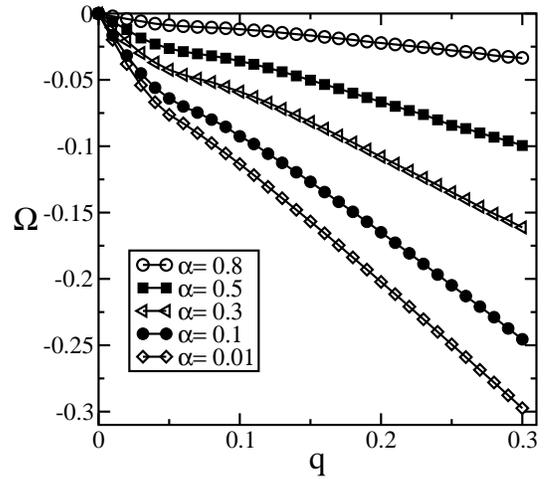}
\caption{Imaginary part of the growth rate $\Omega(q)$ for different values 
of the anisotropy coefficient $\alpha$, with velocity $v= 1.0$.}
\label{fig.imdispersio}
\end{figure}

Furthermore,
since vortices are absent in the positive region, we have to impose
that the density of the perturbation related to the $n^+$ field, and
its derivative in space, have to vanish there.  Similarly we get a
second ODE with constant coefficients by considering that the density
of antivortices is constant at large positive distances.  Taking
again $n^-_{\infty}=1$, we get, for $\zeta\gg 1$:
\begin{equation}
\frac{d^2\delta n^-}{d^2 \zeta}+(v+2iqk)
\frac{d\delta n^-}{d \zeta}-(pq^2+\omega +i\Omega)\delta n^-=0.
\label{asymptequation1}
\end{equation}
In order to satisfy the boundary condition, we must consider the
solution which vanishes exponentially.  The solution of this equation
which does not diverge is of the form
\begin{equation}
\delta n^-= C_1 e^{\bar{\lambda} \zeta}\hspace{.2 cm}, 
Re(\bar{\lambda})<0 .\label{asymptsol1}
\end{equation} 
We applied the shooting method in a 4-dimensional space defined
by the free parameters $D_1(0), D_2(0), \omega$ and $\Omega$, by
integrating backward in the interval $[-L,0]$ and then in
$[0,+\infty[$, looking for solutions of the type
(\ref{asymptsol},\ref{asymptsol1}).

By matching the solutions to the boundary conditions       
\begin{equation}
\begin{array}{ll}
\ds\lim_{\zeta \rightarrow -L}n_1^+=- \frac{dn_0^+}{d\zeta}.
\hspace{0.3 cm}&\ds\lim_{\zeta \rightarrow -L}n_2^+=0,\\
\ds\lim_{\zeta \rightarrow +\infty}n_1^-=0,\hspace{0.3 cm}
&\ds\lim_{\zeta \rightarrow +\infty}n_2^-=0,
\label{bcpert}
\end{array}
\end{equation}
 we then obtain a unique dispersion relation for the real part of the
growth rate $\omega(q)$.

\begin{figure}[b]
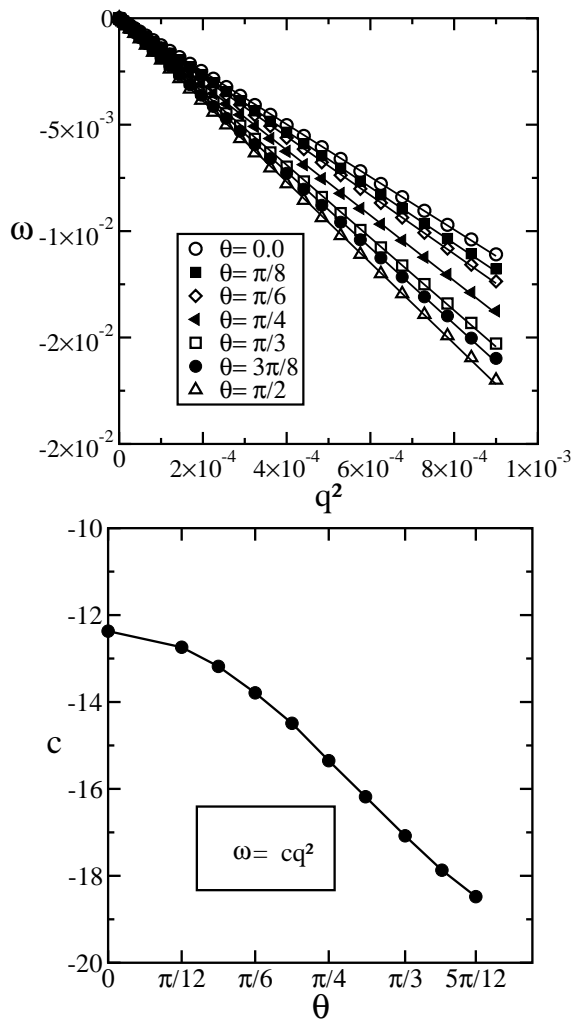

\includegraphics[width= 7.5 cm]{angleregress-1.eps}
\includegraphics[width= 6.5 cm]{regressioresult-1.eps}
\caption{(a) Plot of $\omega(q^2)$ as a function of the angle
 $\theta$. (b) For a coefficient of anisotropy $\alpha = 0.8$ and a
velocity $v= 1.0$, the results from linear regression for the
slope evaluated at $q=0$, $c = d\omega/d(q^2)$, are plotted as a 
function of $\theta$.}
\label{fig.slopealpha}
\end{figure}

\subsection{Results}

Fig.~\ref{fig.aniso} represents the dispersion relation for an angle
$\theta = \pi/4$ and different coefficients of anisotropy $\alpha$.
The front is always stable, even in the presence of very strong
anisotropy, for very low values of $\alpha$.  As the anisotropic
coefficient $\alpha$ is lowered from above, for fixed wavenumber $q$,
the growth rate $\omega (q)$ increases, but it is always negative.
For small $q$ a quadratic behavior of $\omega(q)$ is found:
\begin{equation}
\omega\approx c q^2,\hspace{.2 cm} q\ll 1, \label{omegac}
\end{equation} 
where the (negative) coefficient $c$ depends on the anisotropy of the
sample.  In Fig.~\ref{fig.imdispersio} we have plotted the frequency
$\Omega$ as a function of the wavenumber $q$.  One observes from
(\ref{shiftxi}) that $\Omega/q$ is the velocity with which the
perturbation of the front shifts along the direction transverse to the
propagation direction.  The behavior of $\Omega(q)$ is linear for low
wavenumber $q$ and is proportional to the non-diagonal element of the
mobility tensor $k$,
\begin{equation}
\Omega(q)\propto kq,\hspace{0.2 cm}q \ll 1.
\end{equation}
For an anisotropy coefficient equal to one the isotropic case is
recovered and then $\Omega(q)$ vanishes identically for all
wavenumbers.
As we have already mentioned, the equations that we have used are valid
at scales larger than the cutoff represented by the London penetration 
depth. Anyway, since our results clearly show a stability in the large 
$q$ behavior, our model provides a good description for the dynamics 
of the front.
In Fig.~\ref{fig.slopealpha} we plot the growth rate $\omega$ as a
function of $q^2$ for different values of the angle $\theta$.  Linear
regression then gives a slope corresponding to
the constant $c$ in (\ref{omegac}), which is half the second
derivative of the growth rate $\omega$ with respect the wavenumber at
$q=0$. The dependence of $c$ as a function of the angle $\theta$ is
shown in the lower plot.  As the angle $\theta$ increases, the front
becomes more and more stable.  This behavior can be understood
directly from the form of the equations.  By applying the
transformation
\begin{equation}
\theta \rightarrow \ds{\frac{\pi}{2}}- \theta, \hspace{.2 cm}
0<\theta<\pi/4,
\end{equation}
 the elements of the mobility tensor transform into
\begin{equation}
p\rightarrow {\ds \frac{1}{p}}, \mbox{\hspace{0.5 cm}}k \rightarrow
{\frac{k}{p}}. \label{pqtrans}
\end{equation} 
By considering the quadratic relation of
$\omega(q)$ for small $q$ and the fact that the equations are
invariant under the transformations $\tilde{q}=pq$ and (\ref{pqtrans}),
 it is easy to derive
\begin{equation}
\omega(q)|_{{\theta}} = p^2 \omega(q)|_{{\pi/2 -\theta}},
\hspace{.2 cm} 0<p<1,
\end{equation}
which proves that the dispersion relation becomes more negative as
$\theta$ increases.  When the direction of propagation is that of the
fast growth direction the isotropic case is recovered.
\begin{figure}[t]
\includegraphics[width= 6.0 cm]{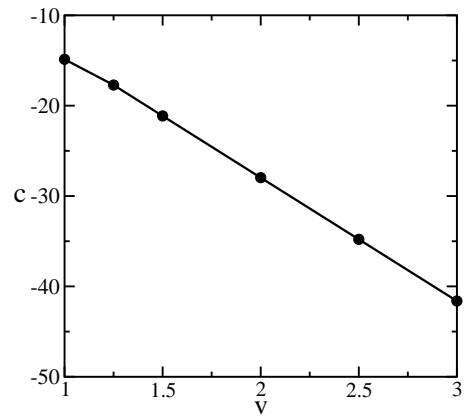}
\caption{Velocity dependence of half the second derivative of
$\omega(q)$ with respect to $q$ evaluated at $q=0$.  As the velocity
increases the front becomes more and more stable.}
\label{fig.veloxdep}
\end{figure}
In Fig.~\ref{fig.veloxdep} we show the dependence of the coefficient
$c$ as a function of the velocity of the front. The front is stable
for velocities for which $n^+_0$ vanishes linearly
($v \geq 1/4$). Furthermore the front becomes more stable with increasing
$v$.  As one can easily understand from the form of the unperturbed
front, the vortex density profile becomes steeper with increasing the
velocity. The limit of infinitely large $v$ corresponds to the case of a front
of vortices propagating in the absence of antivortices.  Thus, the
results confirm the stability of the front without an opposing flux of
antivortices.\\

\section{Stationary front}
{
\label{quattro}
\begin{figure}[t]
\includegraphics[width= 6.0 cm]{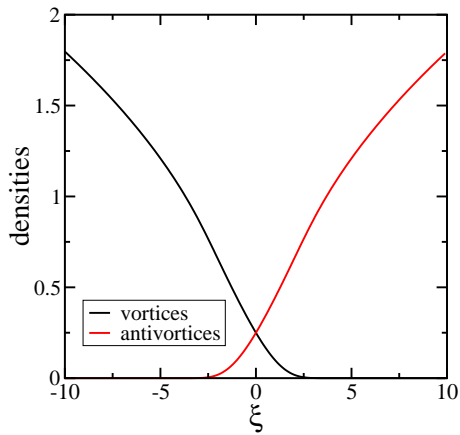}
\caption{Density profiles for vortices and antivortices in the
stationary case ($v=0$). The profiles are smooth and are not
characterised by singularities, as was the case for fronts
propagating with finite velocity.}
\label{fig.stationaryprof}
\end{figure}
As we mentioned in the introduction, we have also analysed the case of
a stationary front, with $v=0$. In this case it is easy to derive the
unperturbed profiles for the densities of vortices and antivortices,
since they are continuous and do not present any singularities. This
case was previously studied in (\cite{Fisher}) and treated in terms of
a sharp interface limit. Equations
(\ref{comovingframe1},\ref{comovingframe2}) in this case simplify
to:
\begin{eqnarray}
\frac{d}{d\xi}n_0^{+}\,\frac{d}{d\xi}(n_0^{+}-n_0^{-})-n_0^{+}
n_0^{-}&=&0,\label{comovingframe3}\\
\frac{d}{d\xi}n_0^{-}\,\frac{d}{d\xi}(n_0^{-}-n_0^{+})-n_0^{+}
n_0^{-}&=&0.\label{comovingframe4}
\end{eqnarray}
The profiles of vortices and antivortices are symmetric in this case,
and outside the interfacial zone the density fields can be easily
derived analytically.  By neglecting the annihilation term,
the profiles of vortices and antivortices have a dependence on the
coordinate $\xi$ of the type:
\begin{equation}
n_0^{\pm}= \sqrt{N^2\mp2C(\xi \pm \xi_1)}, \label{squaredep}
\end{equation} 
where $]-\xi_1,\xi_1[$ denotes the region where vortices and
antivortices overlap, N is the density at $(\pm \xi_1)$ and C a
constant.  The density of vortices and antivortices decays with a
Gaussian tail, as can easily be calculated from equations
(\ref{comovingframe3} and \ref{comovingframe4}). For
Eq.~(\ref{comovingframe3}), by considering that $n_0^+$
assumes a Gaussian-like dependence, and from the form of
(\ref{squaredep}), we get the following equation:
\begin{equation}
-\frac{dn_0^+}{d\xi}\frac{dn_0^-}{d\xi}=n_0^+n_0^-.
\end{equation}
This yields in a self-consistent way a Gaussian behaviour for $n^+$:
\begin{equation}
n_0^+\approx A e^{-\xi^2-\xi(N^2/C-2\xi_1)},
\end{equation} 
where $A$ is a constant.  The density profiles for vortices and anti-vortices are represented in Fig.7. The stability of the front was studied by
following a similar procedure as for the moving front. Because of the
regular profiles, the ansatz (\ref{shiftxi}) that we have applied for
the case of a finite velocity is not required. Thus the linear
stability analysis can be carried out in the standard way and the
linearised equations for the perturbation can easily be integrated.
We do not explain here the procedure in detail, since it is a
simplified version of the one discussed in the previous section.
\begin{figure}[b]
\includegraphics[width= 7.0 cm]{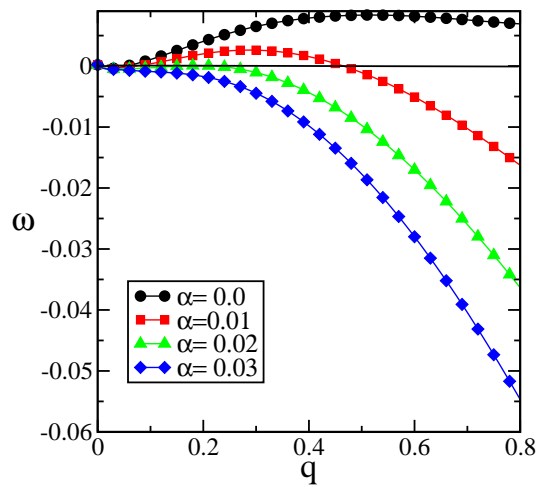}
\caption{Dispersion relation $\omega(q)$ in the case of a stationary
front. An instability is found for a critical anisotropy coefficient
$\alpha_c\approx0.02$.}
\label{fig.stationarydispersio}
\end{figure}
As Fig.~\ref{fig.stationarydispersio} shows, an instability is found
below a critical coefficient of anisotropy $\alpha_c\approx
0.02$. These results confirm previous approximate calculations
\cite{Fisher}, but, as we have already underlined, this coefficient
would correspond to an extremely high in-plane anisotropy which is not
found in any type of superconducting material.  We conclude that this
model of a stationary front in the presence of anisotropy is
insufficient to explain the turbulent behaviour that has been found
experimentally at the flux-antiflux boundary.

}
\section{Conclusions}
{
\label{cinque}

From our analysis it follows that the planar front of vortices moving
with a sufficiently large velocity $v$ in a superconducting thin film
is stable even in the presence of strong in-plane anisotropy. For
stationary fronts, on the other hand, our stability analysis confirms
the earlier approximate analysis of \cite{Fisher}, confirming that
such fronts show an instability to a modulated state in the limit of
very strong anisotropy. From an experimental point of view, the
critical anisotropy of this instability is very high when compared
with real values that can be found for materials with both tetragonal
and orthorhombic structure \cite{anisocoeff2,anisocoeff3}, even when a
non-linear current-electric field characteristic is considered
\cite{Fisher1}.  From a theoretical point of view, the behavior in the
limit of small but finite $v$ is still open as we have not
investigated the range $0<v<1/4$ where the profiles have a noninteger
power law singularity. It could be that the instability gradually
becomes suppressed as $v$ increases from zero, or it could be that the
limit $v\to0$ is singular, and that moving fronts are stable for any
nonzero $v$. Only further study can answer this question.

Our calculations differ markedly from previous work in that we focus on
moving fronts from the start, where our results follow from a
straightforward application of linear stability analysis to our model.
Taken together, these results lead to the conclusion that a model
which includes a realistic in-plane anisotropy, but which neglects the
coupling with the temperature, cannot explain the formation of an
instability at a vortex-antivortex boundary for sufficiently large
front velocities. At the same time, our calculations show that the
issue of the stability of vortex fronts is surprisingly subtle and
rich. For example, we note the fact that for any front velocity, the
value $n^-_0$ at the singular line is exactly $n_\infty/2$ for any
$v$. Is this simply a mathematical curiosity or is the absence of
instabilities related to this unexpected feature through the 
boundary conditions at infinity? Is the presence of a
gradient in the antivortex distribution far ahead of the front
perhaps necessary to generate a long-wavelength front instability?
These are all still open issues, so clearly it is difficult to make
general statements about the (transient) stability of such fronts in
less idealized situations.

One possible interpretation of the results is that when one has a
finite slab into which vortices penetrate from one side, and
antivortices from the other side, a stationary modulated front
(anihilation zone) forms in the middle for extremely large
anisotropies. However, a moving front never has a true
Mullins-Sekerka type instability, since a protrusion of the front
into the region of antivortices is always damped as a result of the
increased annihilation.

The fact that the turbulent behavior at the interface between vortices
of opposite sign was found in a temperature window \cite{Koblischka},
shows that the coupling with the local temperature in the sample has
to be considered.  It appears that it is necessary to include both the
heat transport and dissipation in the model.  Applying an appropriate
stability analysis to such extended models is clearly an important
issue for the future.

\section{Acknowledgment}
We are grateful to Hans de Haan for performing time-dependent
simulations of the one-dimensional vortex dynamics equations, which
showed that we were initially mistaken about the nature of the induced
singularities in the $n^-$ field. In addition we have profited from
discussions with Peter Kes, Rinke Wijngaarden and Gianni Blatter. C.B. is
supported by the Dutch research foundation FOM, and would
also like to thank the ``Fondazione A. Della Riccia'' and ``Fondazione
ing. A. Gini'' for additional financial support. M.H. acknowledges
financial support from The Royal Society.
\appendix
{
\section{Linearized equations for the perturbed front}
From the linear stability analysis we get the linearized equations for 
the variables $D$ and $S$:
\begin{eqnarray}
&&\omega\left(D_1+\frac{dD_0}{d\zeta}\right)-\Omega D_2 +p q^2 S_0
\left(D_1+\frac{dD_0}{d \zeta}\right)=\\
&&+v \frac{dD_1}{d\zeta}+\left(\frac{dS_0}{d\zeta}\right)
\left(\frac{dD_1}{d\zeta}\right)+\left(\frac{dD_0}{d\zeta}\right)
\left(\frac{dS_1}{d\zeta}\right)\nonumber\\
&&+ S_0\frac{d^2D_1}{d \zeta^2}+S_1\frac{d^2D_0}{d\zeta^2}-qk
\left[2S_0\frac{dD_2}{d\zeta}+\frac{dS_0}{d\zeta}D_2+
\frac{dD_0}{d\zeta}S_2\right]\nonumber\\
&&\omega D_2+\Omega \left(D_1+\frac{dD_0}{d\zeta}\right)+pq^2S_0D_2=\\
&&+v \frac{dD_2}{d\zeta}+\left(\frac{dS_0}{d\zeta}\right)\left(\frac{dD_2}{d\zeta}\right)+\left(\frac{dD_0}{d\zeta}\right)\left(\frac{dS_2}{d\zeta}\right)\nonumber\\
&&+ S_0\frac{d^2D_2}{d \zeta^2}+S_2\frac{d^2D_0}{d\zeta^2}+qk\left[2S_0\left(\frac{dD_1}{d\zeta}+\frac{d^2D_0}{d\zeta^2}\right)\right.\nonumber\\
&&+\left.\frac{dS_0}{d\zeta}\left( D_1 + \frac{dD_0}{d\zeta}\right)+\frac{dD_0}{d\zeta}\left(S_1+\frac{dS_0}{d \zeta}\right)\right]\nonumber\\
&&\omega\left(S_1+\frac{dS_0}{d\zeta}\right)-\Omega S_2 +pq^2 D_0\left(D_1+\frac{dD_0}{d \zeta}\right)=\\
&&+v \frac{dS_1}{d\zeta}+2\left(\frac{dD_0}{d\zeta}\right)\left(\frac{dD_1}{d\zeta}\right)+ D_0\frac{d^2D_1}{d \zeta^2}\nonumber\\
&&+D_1\frac{d^2D_0}{d\zeta^2}-qk\left[2D_0\frac{dD_2}{d\zeta}+2\frac{dD_0}{d\zeta}D_2\right]-S_0S_1+D_0D_1\nonumber\\
&&\omega S_2+\Omega \left(S_1+\frac{dS_0}{d\zeta}\right)+pq^2D_0D_2=\\
&&+v \frac{dS_2}{d\zeta}+2\left(\frac{dD_0}{d\zeta}\right)\left(\frac{dD_2}{d\zeta}\right)+ D_0\frac{d^2D_2}{d \zeta^2}+D_2\frac{d^2D_0}{d\zeta^2}\nonumber\\
&&+qk\left[2D_0\left(\frac{dD_1}{d\zeta}+\frac{d^2D_0}{d^2\zeta}\right)+2\frac{dD_0}{d\zeta}\left(D_1+\frac{dD_0}{d\zeta}\right)\right]\nonumber\\
&&-S_0S_2+D_0D_2\nonumber
\end{eqnarray}
} 


\begin{thebibliography}{99}
\bibitem{note3} In
mathematical terms the field $n^+$  has compact support.
\bibitem{note4} To simplify the presentation, we have included only
the terms we need for our discussion. One can generalize the analysis
by  including  regular terms with integer exponents, but the analysis
then shows that these terms have to be absent if $\alpha$ is
non-integer.
\bibitem{note5} Actually, since the jump in the derivative of $n^+$ is
equal to the jump in the derivative of $n^-$ (compare Eq.~(\ref{jump}),
the derivatives of $D_1$ and $D_2$ are unique at $\zeta=0$. The
derivative of $S$ has a jump, however.
\end{thebibliography}
\end{document}